\documentclass[twocolumn]{aastex631}

\usepackage{graphicx}
\usepackage{amsmath}
\usepackage{float}
\usepackage{natbib}
\usepackage{tabularx}
\usepackage{bm}
\usepackage{appendix}
\usepackage{longtable}
\usepackage{enumitem}

\newcommand\msun{M_{\odot}}

\shorttitle{Redshift Evolution of BBHs from GCs}
\shortauthors{Ye \& Fishbach}

\graphicspath{{./}{figures/}}

\begin{document}

\title{The Redshift Evolution of the Binary Black Hole Mass Distribution from Dense Star Clusters}

\author[0000-0001-9582-881X]{Claire S. Ye}
\affil{Canadian Institute for Theoretical Astrophysics, University of Toronto, 60 St. George Street, Toronto, Ontario M5S 3H8, Canada}
\correspondingauthor{Claire S.~Ye}
\email{claireshiye@cita.utoronto.ca}

\author[0000-0002-1980-5293]{Maya Fishbach}
\affil{Canadian Institute for Theoretical Astrophysics, University of Toronto, 60 St. George Street, Toronto, Ontario M5S 3H8, Canada}

\begin{abstract}

Gravitational-wave detectors are unveiling a population of binary black hole (BBH) mergers out to redshifts $z \approx 1$, and are starting to constrain how the BBH population evolves with redshift. We present predictions for the redshift evolution of the BBH mass and spin distributions for systems originating from dense star clusters. Utilizing a grid of 144 state-of-the-art dynamical models for globular clusters, we demonstrate that BBH merger rates peak at higher redshifts for larger black hole primary masses $M_1$. Specifically, for $M_1\gtrsim40\,\msun$, the BBH merger rate reaches its peak at redshift $z\approx2.1$, while for $M_1\lesssim20\,\msun$, the peak occurs at $z\approx1.1$, assuming that the cluster formation rate peaks at $z=2.2$. The average BBH primary mass also increases from $\sim 10\,\msun$ at $z=0$ to $\sim 30\,\msun$ at $z=10$. We show that $\sim 20\%$ BBHs contain massive remnants from next-generation mergers, with this fraction increasing (decreasing) for larger (smaller) primary masses. This difference is not large enough to significantly alter the effective spins of the BBH population originating from globular clusters, and we find that their effective spin distribution does not evolve across cosmic time. These findings can be used to distinguish BBHs from dense star clusters by future gravitational wave observations.

\end{abstract}


\section{Introduction} \label{sec:intro}
Gravitational wave (GW) detections of binary black hole (BBH) mergers by Advanced LIGO \citep{advanceLIGO}, advanced Virgo \citep{advanceVIRGO}, and KAGRA \citep{KAGRA} are rapidly expanding our understanding of the formation and evolution of this population of stellar remnants. To date, 90 GW events have been detected by the LIGO-Virgo-KAGRA collaboration (LVK), including $\sim 80$ BBH mergers \citep{GWTC3}. These detections have sparked searches for their origins, which can offer profound insights into not only the evolutionary processes of massive stars and binaries leading to BBHs but also shed light on their birth environments including dense star clusters \citep{Romero+2021,Fishbach_Fragione_2023,Kou+2024}. 

Various channels have been proposed for the formation of BBHs \citep[see e.g.,][for reviews]{Mapelli_2021,Mandel_Broekgaarden_2022}, which can be broadly divided into two mechanisms: isolated binary evolution \citep[e.g.,][]{Belczynski+2002, Belczynski+2016,Marchant+2016,Heuvel+2017} and dynamical evolution \citep[e.g.,][]{Rodriguez+2016,Antonini+2017,Banerjee_2017,Banerjee_2018,Dicarlo+2019,Arcasedda_2020,Rastello+2021,Ford_McKernan_2022,Arcasedda+2024c}. It has been shown that BBHs formed in isolated binaries and dynamically active star clusters can possess different observable properties including masses, spins, and eccentricities \citep[e.g.,][and references therein]{Mandel_Farmer_2022}.

Limited by statistical uncertainties (due to the small number of detections) and systematical uncertainties (due to imperfect simulations), however, the contributions from different channels are still uncertain, particularly as a function of redshift \citep[e.g.,][]{Zevin+2021,Mapelli+2022, Arcasedda+2023,Cheng+2023,Raikman+2024}. A greater number of detections will impose more stringent constraints on the redshift evolution of properties such as the merger rate \citep[e.g.,][]{Fishbach+2018}, the mass distribution \citep[e.g.,][]{Fishbach+2021_when}, and the spin distribution \citep[e.g.,][]{Biscoveanu+2022}. Furthermore, next-generation GW detectors, such as Einstein Telescope \citep{einstein_telescope}, and Cosmic Explorer \citep{cosmic_explorer}, will be able to probe BBH populations past redshifts $z\gtrsim10$ \citep{Hall_Evans_2019}, providing deeper understanding of the formation mechanisms and evolution environments of BBHs.

Here, we study the evolution of the BBH mass distribution as a function of redshift for BBH mergers from globular clusters (GCs), taking into account cluster formation properties and rates. Previous studies have only briefly discussed how the BH masses vary at selected redshifts \citep[e.g.,][]{Mapelli+2022,Kou+2024,Torniamenti+2024}, employed semi-analytical methods to calculate the redshift evolution of merger rates across different BBH chirp masses \citep{Fujii+2017}, or focused mostly on the merger rate distributions for high-mass BHs \citep{Rodriguez+2016}. We improve upon earlier studies by utilizing a large grid of GC simulations computed with the state-of-the-art \texttt{Cluster Monte Carlo} code (\texttt{CMC}; \citealp[e.g.,][and references therein]{CMC1}) that incorporates various relevant physics for cluster dynamics and the evolution of the stars and compact objects within. We also investigate the detailed characteristics of the redshift evolution of merger rates and masses across BBH mass ranges.

The paper is organized as follows. We describe our methods for calculating the BBH merger rates as a function of redshift from GCs in Section~\ref{sec:gcs}. In Section~\ref{sec:bbhs}, we present the different merger rate distributions corresponding to various BH mass ranges and the BH mass evolution as a function of redshift. We also explore the effective spin distributions as a function of redshift. We discuss the uncertainties of the cluster simulations in Section~\ref{sec:uncer}. Finally, we summarize and conclude in Section~\ref{sec:conclu}.

\section{Globular Cluster Populations}\label{sec:gcs}
\subsection{Globular Cluster Simulations}\label{subsec:catalog}
We use a large catalog of 144 GC simulations \citep{Kremer+2020catalog} ran with \texttt{CMC} to explore the mass distribution as a function of redshift for merging BBHs from GCs. These catalog models span a wide range of initial conditions, with initial number of stars $N = 2\times10^5$, $4\times10^5$, $8\times10^5$, and $1.6\times10^6$ (corresponding to initial masses of about $1.2\times10^5$, $2.4\times10^5$, $4.8\times10^5$, and $9.7\times10^5\,\msun$, respectively), initial virial radius $r_v = 0.5$, 1, 2, and $4\,$pc, metallicity $Z =$ 0.0002, 0.002, and 0.02, and galactocentric distance $r_g = 2$, 8, and $20\,\rm{kpc}$. The galactocentric distance sets the tidal boundaries of the clusters in their host galaxies. All simulations adopt an initial mass function following the standard Kroupa broken power-laws \citep{Kroupa_2001} in the mass range $0.08-150\,\msun$. The simulations use a King profile \citep{King1966} with a concentration parameter $W_0=5$ for the stellar distribution. The initial binary fraction is $5\%$ for all primary masses, and the secondary masses are sampled from a uniform mass ratio between 0.1 and 1 of the primary masses \citep{Duquennoy_Mayor_1991}. Binary orbital periods are drawn from a distribution flat in log-scale \citep[e.g.,][]{Duquennoy_Mayor_1991} ranging from near-contact ($\geq5(R_1+R_2)$, where $R_1$ and $R_2$ are the radii of the binary component stars) to the hard/soft boundary, and the binary eccentricities are drawn from a thermal distribution \citep[e.g.,][]{Heggie_1975}. The present-day properties of these models match well with those of the observed Milky Way GCs \citep[][their Figure~2]{Kremer+2020catalog}.

The simulations adopt the ``rapid model" for stellar remnant formation from core-collapse supernova \citep{Fryer+2012}. Specifically, BHs form with mass fallback. Their natal kicks are sampled from a Maxwellian distribution with velocity dispersion of $\sigma = 265\,{\rm km\,s^{-1}}$ \citep{Hobbs+2005}, but with reduced magnitude by a factor of $1-f_{\rm fb}$, where $f_{\rm fb}$ is a parameter measuring the mass fraction of the stellar envelope that falls back upon core collapse.

Pulsational-pair instabilities and pair-instability supernovae are also incorporated into the simulations, following the treatment in \citet{Belczynski+2016}. Massive stars with pre-explosion helium core masses between $45$ and $65\,\msun$ undergo pulsations that eject large amounts of mass, leading to a final stellar mass of at most $45\,\msun$. The collapse of the final products produce BHs of $40.5\,\msun$ (assuming $10\%$ mass loss converting from baryonic to gravitational matter). Stars with helium core masses above $65\,\msun$ are destroyed by pair-instability supernovae leaving behind no remnants.

We assume that all BHs from stellar collapse are born with zero spin. New spins and recoil kicks are applied to the remnants immediately after the BBH mergers, allowing self-consistent computations of BH retention by the model clusters \citep[][and references therein]{Rodriguez+2018_PNspin}. \texttt{CMC} includes post-Newtonian dynamics for BHs \citep{Antognini+2014,Amaro_Chen_2016,Rodriguez+2018_PNeccentric,Rodriguez+2018_PNspin} and models GW emission and BBH mergers during strong encounters self-consistently.

\begin{figure}
\begin{center}
\includegraphics[width=\columnwidth]{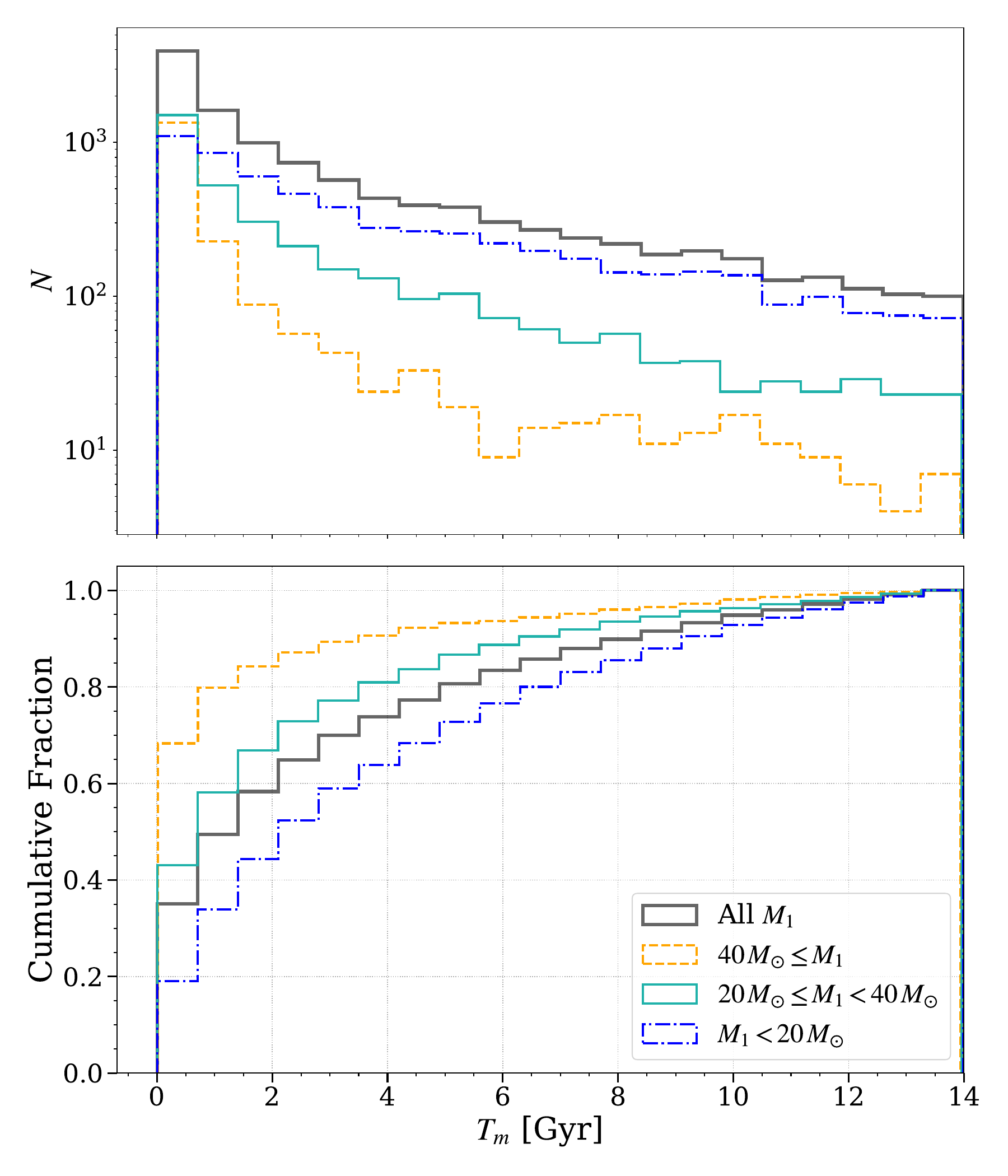}
\caption{The distributions of BBH merger times since the formation of their host clusters from the catalog models. Four different primary mass ($M_1$) ranges are shown. About $80\%$ of BBHs with $M_1\geq40\,\msun$ merged within the first Gyr of the evolution of their host clusters, while less than $40\%$ of BBHs with $M_1<20\,\msun$ merged in the same period.}\label{fig:mass_time}
\end{center}
\end{figure}

Figure~\ref{fig:mass_time} shows the merger time distributions of various primary mass ranges for all BBHs that merged within a Hubble time from the catalog models. For this plot, we weight all clusters in the \texttt{CMC} catalog equally. Here, $t=0$ corresponds to the beginning of the cluster evolution. We find that BBHs with larger primary masses tend to merge at earlier times. In particular, most ($\sim 80\%$) BBHs with $M_1 \gtrsim 40\,\msun$ merged within one Gyr after their host clusters are formed, while only $\sim 30\%$ of BBHs with $M_1 \lesssim 20\,\msun$ merged in the first Gyr. This is because massive BHs form early in the first few Myrs of stellar evolution. They also mass segregate to the dense cluster cores on similar timescales \citep[][their Eq. 20]{PZ+2010}, where frequent gravitational encounters allow them to form binaries and merge at early times. If we approximate the delay time distribution with a power law $p(\tau) \propto \tau^{\alpha}$, the best-fit power-law indexes are $\alpha \approx -1, -0.8, -1.2$, and $-1.4$ for all primary masses, $M_1<20\,\msun$, $20\,\msun \leq M_1<40\,\msun$, and $M_1 \geq 40\,\msun$, respectively.

\subsection{Cluster Formation Rate}\label{subsec:formation}
The population properties of BBHs as a function of redshift depend on their host clusters' initial properties and formation times. We sample GCs according to their initial mass, virial radius, and metallicity.

We assume that the initial cluster mass distribution follows a Schechter function between $10^5$ and $10^6\,\msun$, similar to the present-day masses of Milky Way GCs and corresponding to the mass span of the catalog models. The mass distribution is described by 
\begin{equation}
    p(M_{\rm GC}) \propto \left(\frac{M_{\rm GC}}{M_S}\right)^{\beta_S} {\rm exp} \left(-\frac{M_{\rm GC}}{M_S}\right)\,,
\end{equation}
where $\beta_S = -2$ and the Schechter mass $M_S = 10^{6.26}\,\msun$ \citep{PZ+2010,Antonini_Gieles_2020}. 

For the initial size distribution, we adopt a Gaussian distribution following \citet{Fishbach_Fragione_2023}
\begin{equation}
    p(r_v) \propto {\rm exp} \left(-\frac{(r_v - \mu_r)^2}{2\sigma_r^2} \right)\,,
\end{equation}
with mean $\mu_r=2~$pc and standard deviation $\sigma_r=2~$pc. We truncate the distribution at 0.5~pc and 4~pc \footnote{Varying the mean and standard deviation of the initial size distribution within this range does not change the trends of the merger rate peaks and slopes shown later.}.

We assume a lognormal metallicity distribution at each formation redshift
\begin{equation}
    p(Z) \propto {\rm exp} \left(-\frac{({\rm log}_{10}Z - \mu_Z)^2}{2\sigma_Z^2}\right)\,,
\end{equation}
where $\sigma_Z = 0.5$ and $\mu_Z$ is given by the following redshift ($z$) dependent equation from \citet{Madau_Fragos_2017},

\begin{equation}
    {\rm log}_{10} \langle Z/Z_{\msun} \rangle = 0.153-0.074z^{1.34}\,.
\end{equation}

For each redshift, we adopt a cluster formation rate that follows the star formation rate \citep{Madau_Fragos_2017} and is described by 
\begin{equation}\label{eq:form_rate}
    \mathcal{R}_{\rm GC}(z) = \mathcal{R}_0 \frac{(1+z)^{a_z}}{1+[(1+z)/(1+z_{peak})]^{a_z+b_z}}\,,
\end{equation}
where $a_z = 2.6$, $b_z=3.6$, and $z_{peak} = 2.2$ is the peak cluster formation redshift. The maximum cluster formation redshift is set to 20, beyond which $\mathcal{R}_{\rm GC} = 0$. Note that while \citet{Fishbach_Fragione_2023} has inferred from GW observations that the rate of GC formation may rise more sharply at $z>3$ compared to the global star formation rate, simulations of cluster formation indicate that the peak redshift for cluster formation is consistent with being at $z \sim 2$ \citep[e.g.,][their Figure~10]{Reinacampos+2022}. Thus, for simplicity, we assume that cluster formation follows the star formation rate.

The cluster formation rate is normalized by the number density of GCs observed at the present day. We define the total number density of dense clusters ever formed as
\begin{equation}
    n_0 = \int_{0}^{t_{\rm max}} \mathcal{R}_{\rm GC}(t)dt\,, 
\end{equation}
where $t_{\rm max}$ is the maximum lookback time corresponding to a maximum redshift of 20. Many clusters that formed have dissolved or disrupted due to effects including stellar evolution and tidal stripping by their host galaxies. The total number density of dense star clusters formed can be an order of magnitude larger than what we see today \citep[e.g.,][]{Antonini_Gieles_2020}. To take into account the disrupted clusters, we assume $n_0=f_{\rm disrp} n_t$, where $f_{\rm disrp}$ is the disruption factor and $n_t$ is the number density of GCs observed. We adopt $n_t=2.31\,{\rm Mpc^{-3}}$ \citep[][and references therein]{Rodriguez+2016} and adjust the disruption factor such that the total BBH merger rate from GCs at redshift $z=0$ is $10\,{\rm Gpc^{-3}\,yr^{-1}}$ (since the focus of this study is not on the exact amplitude of the BBH merger rate). In this case, $f_{\rm disrp} \sim 5$, consistent with \citet{Antonini_Gieles_2020}. Note that this disruption factor corresponds to a minimum cluster mass $\sim 10^5\,\msun$ for the catalog models. It will likely be larger if lower mass clusters (e.g., $\sim 10^4\,\msun$) are taken into account since they produce smaller numbers of BBHs \citep[e.g.,][]{Rodriguez_Loeb_2018, Antonini_Gieles_2020_popsyn} and dominate the number of clusters formed.

\section{Binary Black Hole Mergers from Globular Clusters}\label{sec:bbhs}
\subsection{Merger Rates and Masses}\label{subsec:merger_rate}

The BBH merger rate for a cluster is calculated by summing over all BBH mergers from the cluster following the method in \citet{Fishbach_Fragione_2023},
\begin{multline}
    \mathcal{R_{\rm BBH}}(z_l) = p(M_{\rm GC}^i) M_{\rm GC}^i p(r_v^j) r_v^j\\
    \times \sum_n \mathcal{R_{\rm GC}}(\hat{z}(t_l+t_m)) p(Z^k|\hat{z}(t_l+t_m)) Z^k\,.
\end{multline}
We weigh each catalog model with its initial mass $M_{\rm GC}^i$, virial radius $r_v^j$, and metallicity $Z^k$ using the distribution functions and cluster formation rate described in Section~\ref{subsec:formation}. Here $t_l$ is the lookback time, and $t_m$ is the time a BBH takes to merge in a simulation. $\hat{z}$ is the cosmological function converting time to redshift, and $\hat{z}(t_l+t_m)$ is the formation redshift of a cluster. We adopt the cosmological parameters in \citet{Planck} as implemented in \citet{Astropy} for the cosmological calculations. The total BBH merger rates from GCs as a function of redshift can be estimated by summing over all catalog models.

\begin{figure}
\begin{center}
\includegraphics[width=\columnwidth]{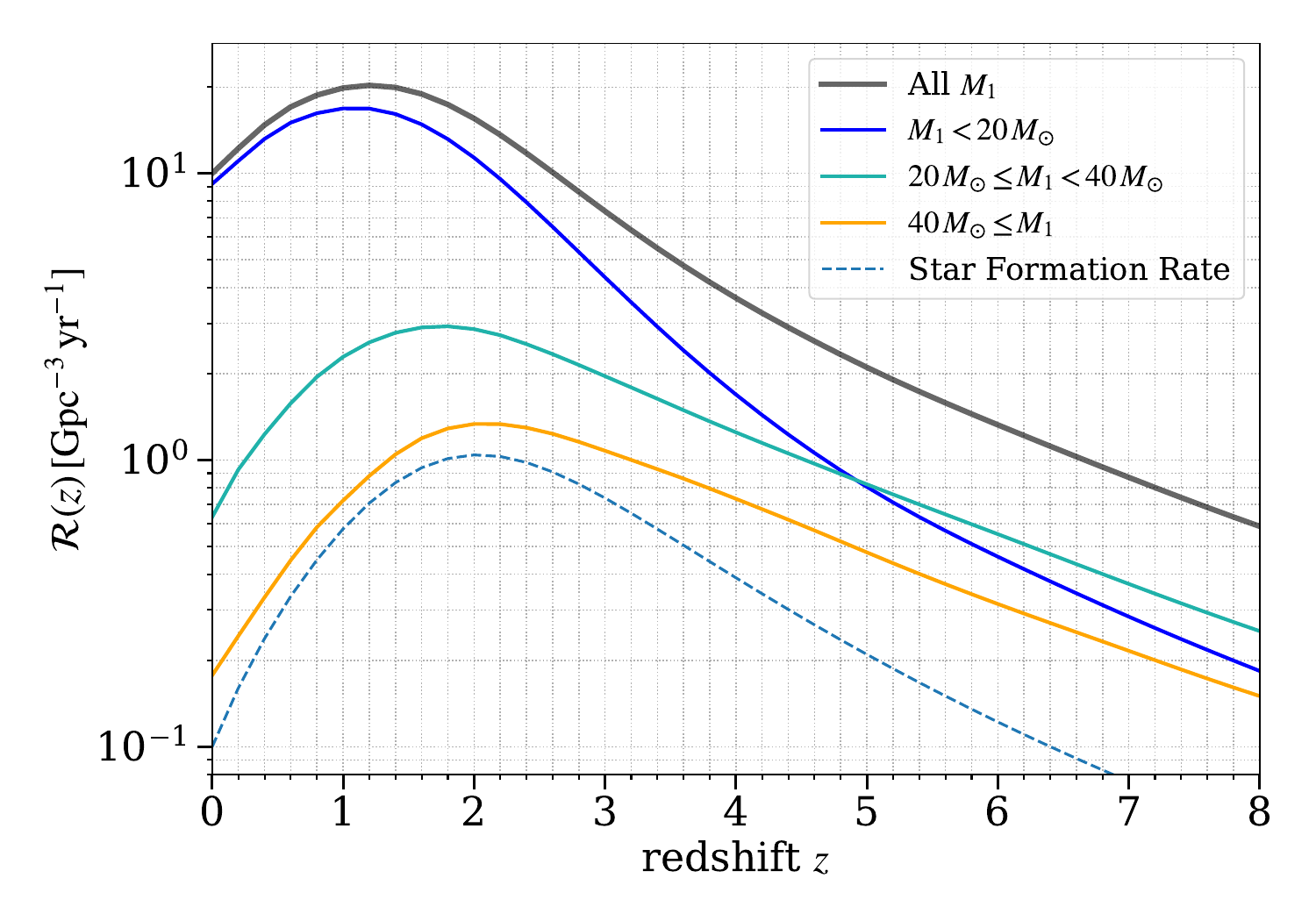}
\caption{BBH merger rates from GCs as a function of redshift for four different primary mass ranges. Mergers with larger primary masses have merger rates peaking at earlier times. The peak merger redshifts are 1.2, 1.1, 1.7, and 2.1 with uncertainties of 0.1 for all BBHs, $M_1<20\,\msun$, $20\,\msun \leq M_1<40\,\msun$, and $M_1 \geq 40\,\msun$, respectively. The merger rate for all BBHs is normalized to $10~{\rm Gpc^{-3}\,yr^{-1}}$ at $z=0$ (see also Section~\ref{subsec:formation}). The cluster formation rate from \citet{Madau_Fragos_2017} is also plotted as the dashed curve (arbitrary normalization), with a peak at $z=2.2$.}\label{fig:R_z}
\end{center}
\end{figure}

Figure~\ref{fig:R_z} shows the BBH merger rates as a function of redshift after taking into account the cluster formation histories. Similar to Figure~\ref{fig:mass_time}, more massive BHs have their merger rates peak earlier due to mass segregation in GCs and the earlier formation time of massive BHs. In addition, the low metallicity environment at higher redshift also facilitates the formation of more massive BHs \cite[e.g.,][]{Torniamenti+2024}. The merger rates peak at redshift about 1.1, 1.7, and 2.1 for BBHs with primary masses in the ranges $M_1<20\,\msun$, $20\,\msun\leq M_1<40\,\msun$, and $M_1 \geq 40\,\msun$, respectively. The merger rate peak of the most massive BHs follows closely the peak of the star and cluster formation rate ($z=2.2$) as described by Eq.~\ref{eq:form_rate}. 

The trend of the peak merger redshifts for different BBH primary masses is distinct from that found in BBH mergers from isolated binary evolution. Recently, \citet{vanSon+2022} showed that common envelope evolution and stable Roche-lobe overflow in isolated binary systems produce BBH mergers with different delay times and masses. BBHs with small primary masses ($\lesssim30\,\msun$) from the common envelope channel dominate the merger rates at high redshift, while BBHs with large primary masses ($\gtrsim 30\,\msun$) from the stable Roche-lobe overflow channel contribute to a large fraction of mergers at low redshift. The combined effects of these two binary evolution processes lead to BBHs with small primary masses merging at higher redshifts than BBHs with high primary masses \citep[][their Figure~8]{vanSon+2022}. In particular, they showed that the peak of the merger rates for BBHs with small primary masses $M_1 \lesssim 10\,\msun$ coincide with the peak of their adopted star formation rate ($z=2.7$), while the merger rate for BBHs with $M_1\gtrsim\,30\msun$ peak at a lower redshift $z=1.9$.

This is opposite to what we find for BBH mergers produced in GCs as is demonstrated in Figure~\ref{fig:R_z}. In the redshift range $z \lesssim 1$ (corresponding to the redshift range of current LVK observations; e.g. ~\citealt{Fishbach_vanSon_2023}), the merger rates shown in Figure~\ref{fig:R_z} can be approximated as power-laws $\mathcal{R}_{\rm BBH} \propto (1+z)^{\kappa}$. We find that $\kappa \approx 1.0$, $0.9$, $1.8$, and $2.1$ for all primary masses, $M_1<20\,\msun$, $20\,\msun \leq M_1<40\,\msun$, and $M_1 \geq 40\,\msun$, respectively. (The $1\sigma$ uncertainties are $\lesssim 0.04$ for all values.) These slopes are all slightly shallower than inferred from the third GW transient catalog. Assuming a nonevolving mass distribution across redshift, \citet{Abbott+2023_population} inferred a power-law slope $\kappa = 2.9^{+1.7}_{-1.8}$ at 90\% credibility (see their Figure~13). This may indicate that the GC formation rate peaks at a higher redshift than assumed here \citep{Fishbach_Fragione_2023}.

However, our focus in this work is on the variation of the slopes between different BBH primary masses. We predict the power-law index governing the redshift evolution to vary by $\approx 1$ across the mass range. This is too small to be resolvable with the current GW data, which can only constrain the overall merger rate evolution to a 90\% credibility width of $\approx 3$. Approximately 500 BBH events with the design sensitivity of Advanced LIGO will tighten the 90\% credibility width to 1~\citep{Fishbach+2018,Fishbach_Kalogera_2021}, so we anticipate that $\mathcal{O}(1000)$ events will provide the necessary resolution to identify the predicted variation with BBH primary mass. This will probably require $\mathcal{O}(1)$ year of observation with A+ sensitivity \citep{Kiendrebeogo+2023}.

\begin{figure}
\begin{center}
\includegraphics[width=\columnwidth]{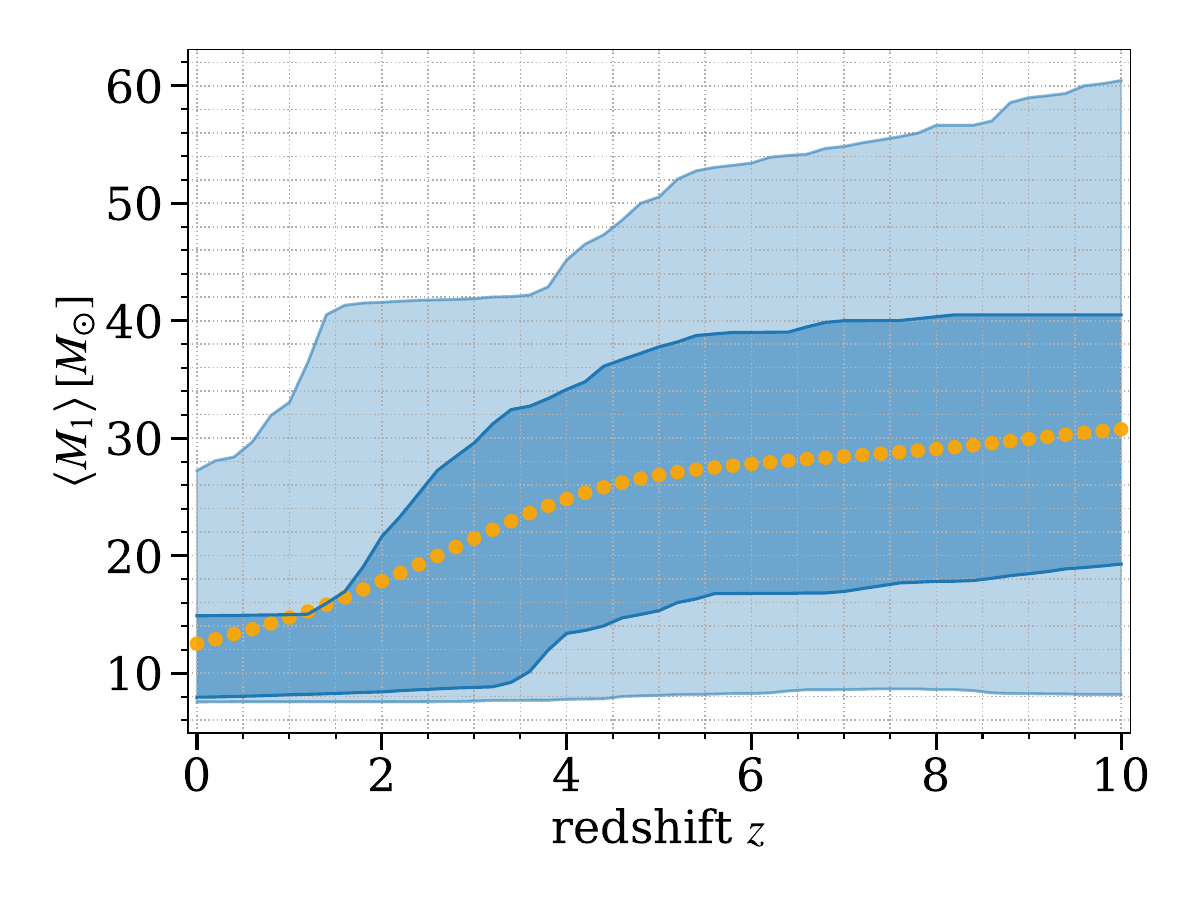}
\caption{The evolution of mass as a function of redshift for BBH mergers from GCs. The average primary mass is shown by the dotted orange curve. The dark blue band shows the 25th and the 75th percentiles of the primary mass, while the light blue band shows the 5th and the 95th percentiles.}\label{fig:avem_z}
\end{center}
\end{figure}

We also show the redshift evolution of the primary mass of the BBH population from GCs in Figure~\ref{fig:avem_z}. The average primary mass increases for larger redshifts, with $\langle M_1 \rangle$ reaching about $30\,\msun$ at $z \approx 10$ and $\langle M_1 \rangle \approx 12\,\msun$ at $z \approx 0$. The average mass of BBH mergers at $z=0$ is about a factor of two smaller than that estimated by \citet{Belczynski+2022}, which may be because they used GC models with low metallicities only (stellar winds are weaker at lower metallicities, thus leading to smaller mass loss and more massive BHs; e.g., \citealp{Belczynski+2010}) and they have a smaller number of models.

Recently, \citet{Rinaldi+2023} suggested that there is evidence for the evolution of BBH primary mass with redshift from the third GW transient catalog \citep{GWTC3}, where the merger rates of BBHs consisting of primary masses $\lesssim 20\,\msun$ drop at $z\gtrsim0.4$. This may indicate a prevalent contribution of dynamically-assembled BBHs at higher redshift. However, \citet{Abbott+2023_population} and \citet{Fishbach+2021_when} showed that within current statistical uncertainties, the GW data is consistent with a nonevolving mass distribution, which matches our expectations that the predicted trend from GCs is not yet resolvable, but will be with future GW observations.

\subsection{Next-generation Mergers and Spins}\label{subsec:high_gen}
In dense star clusters, BBH merger remnants may be retained if the GW recoil velocity is smaller than the escape velocity of the cluster, creating a new generation of BHs. These second-generation (2G) BHs will continue to participate in close dynamical encounters, potentially forming new BBHs that can merge. Therefore, the masses of the merging BBHs may be affected by previous BBH mergers. We define the first-generation BHs (1G) as those formed from collapsing stars and higher-generation BHs as those produced in one or more previous BBH mergers.

\begin{figure}
\begin{center}
\includegraphics[width=\columnwidth]{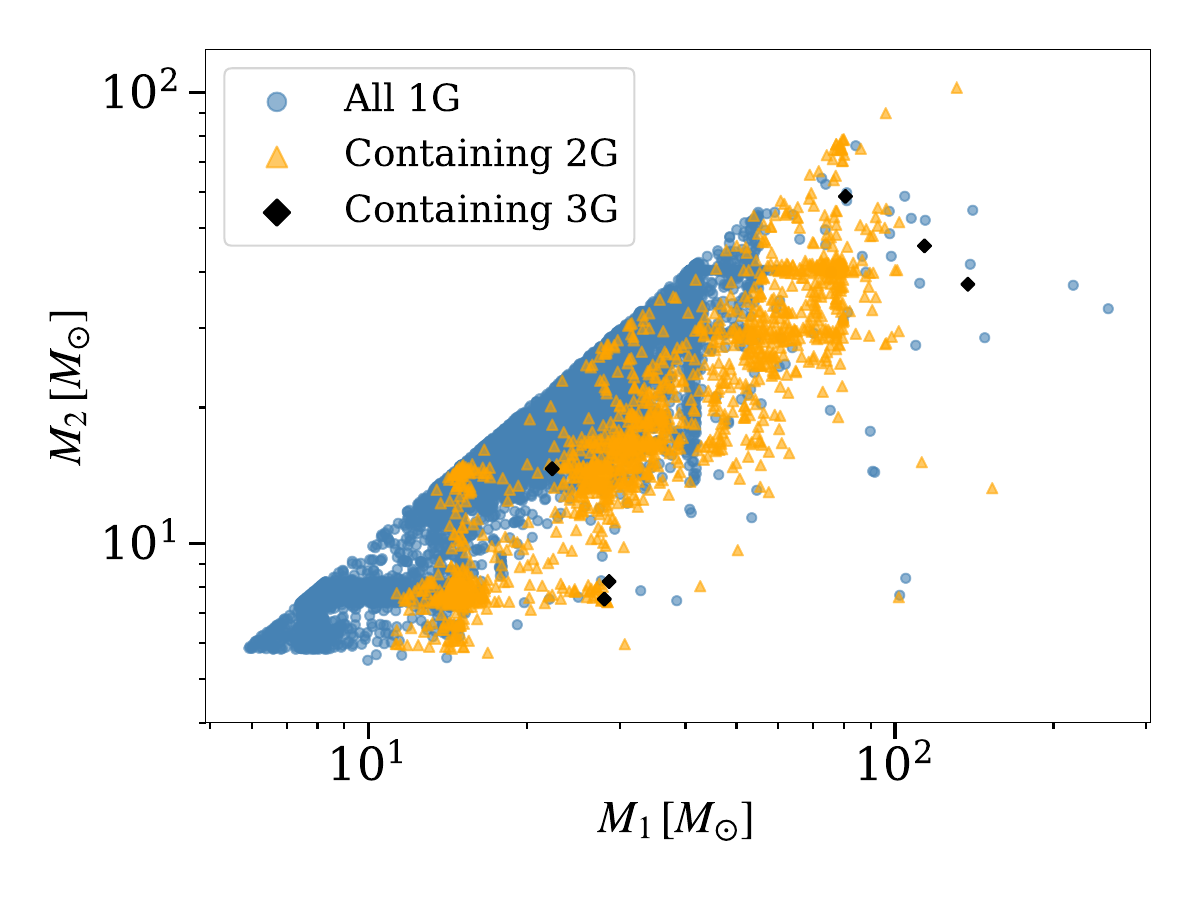}
\caption{Primary and secondary masses of merging BBHs containing different generations of BHs from the catalog models. There are a total of 9517 BBHs that only contain 1G BHs, 1709 that have at least one 2G BH, and 6 that have at least one 3G BH.}\label{fig:m_gen}
\end{center}
\end{figure}

Figure~\ref{fig:m_gen} shows the primary and secondary masses for BBHs containing different generations of BHs from the catalog models. 1G here means both component BHs in a binary are first-generation, while 2G and 3G indicate that at least one BH component is second- or third-generation. Overall, 2G BBHs are more massive than 1G as they contain BH remnants from previous mergers. For primary BH masses in the ranges $M_1<20\,\msun$, $20\,\msun \leq M_1<40\,\msun$, and $M_1 \geq 40\,\msun$, there are about $88\%$, $83\%$, and $70\%$ of binaries that consist of only 1G BHs, respectively.

The fractions of BBHs containing next-generation BHs may affect the population's effective inspiral spins as measured with GWs,
\begin{equation}\label{eq:chieff}
    \chi_{\rm eff} = \frac{\vec{\chi_1}+q \vec{\chi_2}}{1+q}\hat{L}\,,
\end{equation}
where $\vec{\chi_1}$ and $\vec{\chi_2}$ are the spin vectors of the BH components, $q$ is the ratio between the secondary and the primary BH masses, and $\hat{L}$ is the unit binary angular momentum vector. The 2G BHs created by previous mergers of nonspinning BHs have high spins concentrated at $\sim 0.7$ \citep[e.g.,][]{Rodriguez+2019_nextg}, contributing to large effective spins. 

\begin{figure}
\begin{center}
\includegraphics[width=\columnwidth]{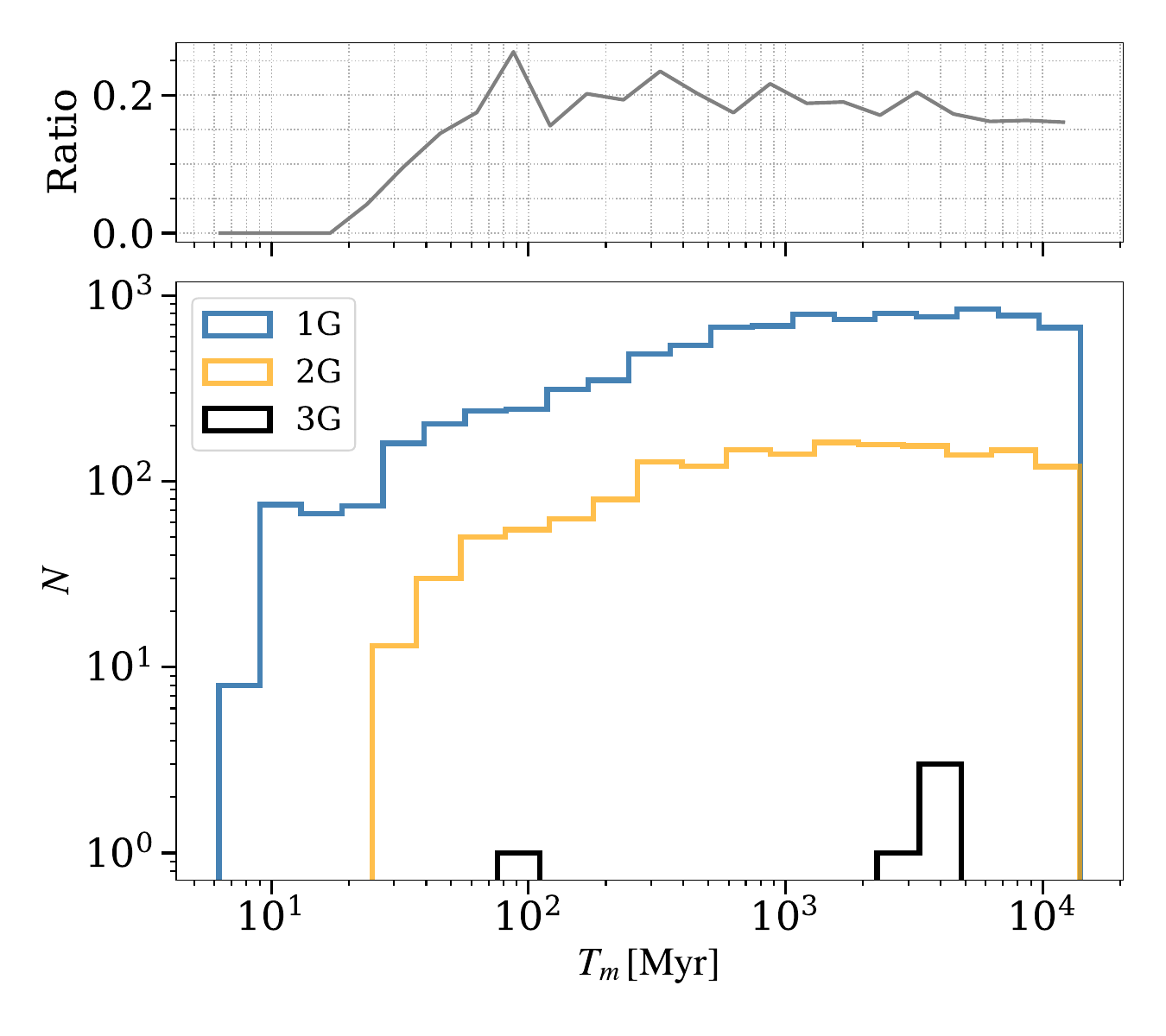}
\caption{Upper panel: the number ratio of next-generation BBH mergers over 1G mergers as a function of merger time since the formation of their host clusters. The ratio stays mostly constant at around 0.2 across a Hubble time except at $\lesssim40$~Myr when the 2G BBHs have not yet formed. Bottom panel: BBH merger time distributions for BBHs containing three different generations of BHs.}\label{fig:ratio_t_gen}
\end{center}
\end{figure}

To further explore if the BBH effective spins also evolve with redshift, we plot the merger time distributions of different BBH generations and the number ratio of next-generation mergers to 1G mergers in Figure~\ref{fig:ratio_t_gen}. The merger time distribution of 2G BBHs is similar to that of 1G BBHs, except there are no 2G mergers at the very beginning ($\lesssim20$~Myr) of the cluster evolution. The merger timescales of 2G BBHs with $M_1\gtrsim\,40\msun$ are shorter compared to those of the 1G BBHs with $M_1\lesssim\,40\msun$, whether inside the clusters or following ejection. Thus, despite the slight delay in the merger times of 2G systems compared to their 1G counterparts, the more massive BBHs still merge at higher redshift, as is demonstrated in Figure~\ref{fig:R_z}. Overall, the number ratio between 1G and higher-generation mergers stays mostly constant at around 0.2 throughout a Hubble time (consistent with \citealp[e.g.,][]{Rodriguez+2019_nextg}) and may only start to decline at $\gtrsim 5$~Gyr after many BHs have been dynamically processed and ejected out of the clusters.

This indicates that the effective spin distribution of the BBH population from GCs remain mostly unchanged across redshift (assuming BHs are born with zero spins from the collapse of stars). This is distinct from the trend predicted by isolated binary evolution, where it is shown that BBH effective spins tend to increase with increasing redshift due to the tidal spin-up of BBH progenitors at high redshift \citep[e.g.,][]{Bavera+2022}. \citet{Biscoveanu+2022} showed that observationally, the width of the effective spin distribution may broaden with increasing redshift, but found no evolution of the mean effective spin with redshift. Thus BBHs from isolated binary evolution or GCs alone cannot explain this observed trend if BHs are born with zero spin from collapsing stars.

In addition, the effective spins also depend on the mass ratios of the component BHs (Eq.~\ref{eq:chieff}). Most BBH mergers from GCs have mass ratios close to unity regardless of their primary masses ($> 50\%$ have $q \geq 0.7$ for all $M_1$ ranges) since mass segregation in dense star clusters leads to the pair-up of similar mass objects. Small mass ratios ($q \lesssim 0.4$; $\sim 10\%$ of all model BBHs) are mostly contributed by BBHs with primary mass $M_1>20\,\msun$, and the most extreme mass ratios ($q\lesssim 0.2$; only $\sim 1\%$ of all model BBHs) are produced by BBHs with $M_1>40\,\msun$. The limited number of massive BHs, in contrast to the abundance of smaller BHs, often results in the pairing of massive BHs with companions of unequal mass, with the minimum mass ratio constrained by the minimum and maximum BH masses in the models. Nevertheless, the small fraction of BBHs with $q \lesssim 0.4$ in the catalog models makes it improbable for these asymmetric systems to significantly affect the effective spins of the overall cluster BBH population.

Only six merging BBHs contain 3G BHs, since 2G BHs created by the previous mergers of two BHs have high spins, and mergers with these fast-spinning 2G BHs produce large GW recoil kicks that will most likely eject the merger remnants from the host clusters \citep{Rodriguez+2019_nextg}. Note that the very massive 1G BHs in the models ($M_{\rm BH} \gtrsim 100\,\msun$) are formed by the collapse of massive stars produced in repeated stellar collisions.

\section{Model Uncertainties and Discussion}\label{sec:uncer}
The properties of the BBHs from GCs could be affected by the adopted initial conditions of the clusters and the assumed BH birth spin. 

For example, the initial binary fractions of cluster stars are not well constrained. Many GCs at the present day are observed to have low binary fractions $\sim 5-10\%$ \citep{Milone+2012}. Young massive star clusters in the local universe, on the other hand, are observed to have high binary fractions comparable to the Galactic field \citep[e.g.,][]{Sana+2009}. In particular, observations have suggested that close to $100\%$ of the O- and B-type stars in the field are born with companions \citep[e.g.,][]{Sana+2012,Moe_DiStefano_2017}. Since it is believed that the progenitors of GCs may have similar properties to the young massive star clusters \citep[e.g.,][]{Chatterjee+2010, Chatterjee+2013b}, massive stars in GCs may have high binary fractions initially. This initial high binary fraction of massive stars can enhance the formation of massive BHs \citep{Arcasedda+2024a} and their mergers \citep{Gonzalez+2021}, in comparison to the catalog models with $5\%$ initial binary fraction for all stars. A larger population of primodial binaries for massive stars may also produce more low-mass BBH mergers (chirp mass $\lesssim20\,\msun$) at short delay times \citep{Hong+2018}, moving the peak redshift of these mergers (e.g., the peak redshift of $M_1<20\,\msun$ in Figure~\ref{fig:R_z}) to higher redshift. However, primordial binaries are unlikely to affect the massive BBH mergers \citep[e.g.,][their Figure~9]{Hong+2018}. Furthermore, \citet{Chatterjee+2017} showed that primordial binaries containing BH progenitor stars are generally disrupted in GCs due to dynamical encounters, producing BBHs independent of the initial binary properties. Therefore, for dense star clusters like GCs, dynamically-produced BBH mergers probably dominate the merger rates regardless of the initial binary fraction, and the initial binaries may have negligible effects on the mass evolution of BBHs mergers in GCs \citep{Hong+2018,Arcasedda+2024a}, preserving the peak redshift trend as a function of the primary BH mass (Figue~\ref{fig:R_z}).

The stellar initial mass function can also influence the populations of BBHs originating from GCs. \citet{Weathreford+2021} showed that top-heavy initial mass functions (those with higher fractions of massive stars) produce more BBH mergers \citep{Chatterjee+2017}, including sources with high BH component masses compared to the canonical initial mass function we adopted \citep[also][]{Wang+2021}.

In addition, the birth spins of BHs formed in stellar collapse determine the GW recoil kicks of BBH mergers and the fractions of 2G BHs that can be retained in GCs. Thus, different BH birth spins will also lead to different mass distributions and effective spin distributions of the BBH population from GCs \citep{Rodriguez+2019_nextg,Mapelli+2021}.

The shapes of the merger rate curves and the absolute merger rates may vary for BBH populations from GCs under different initial conditions and BH birth spin assumptions. However, they are unlikely to alter the trend of the merger rate peaks and slopes as a function of the primary BH mass, as massive BHs will continue to form and undergo mass segregation first.

\section{Conclusions}\label{sec:conclu}
We studied the evolution of mass with redshift for BBH mergers produced in GCs. We calculated the BBH merger rates across different primary mass ranges by utilizing a large grid of realistic GC simulations and taking into account both cluster formation rates and properties. 

Due to mass segregation in dense star clusters and the early formation of massive BHs, BBHs with massive components tend to merge early in the evolutionary history of their host clusters. The low metallicity environment at high redshift also contributes to the formation of massive BHs. We demonstrated that BBHs with primary masses $M_1 \gtrsim 40\,\msun$ have the peak merger rate at $z\approx2.1$, while BBHs with $M_1 \lesssim 20\,\msun$ have the peak merger rate at $z\approx1.1$, assuming a cluster formation rate that peaks at $z=2.2$. Similarly, the average primary mass increases from $\sim 10\,\msun$ at $z=0$ to $\sim 30\,\msun$ at $z=10$. Furthermore, we showed that the power-law index $\kappa$ of the BBH rate evolution at $z \lesssim 1$, presuming $\mathcal{R}(z)\sim (1+z)^{\kappa}$, increases for more massive primary masses from $\kappa \approx 1$ to $\kappa \approx 2$. 

The correlation between the redshift of the merger rate peaks and the primary mass of the merging BBHs from GCs is the opposite of that found in isolated binary evolution \citep{vanSon+2022}. Thus, the redshift evolution of the BBH mass distribution can be used to distinguish their formation channels. Depending on the primary mass spectrum, current GW detections are consistent with a nonevolving mass spectrum but do not yet provide the resolution to probe the degree of mass evolution predicted here \citep{Fishbach+2021_when}. However, we anticipate that during the next couple of observing runs by the LVK, $\mathcal{O}(1000)$ BBH detections at $z \lesssim 2$ will provide the constraining power to test our predictions for the mass-dependence of the merger rate evolution. Furthermore, next-generation GW detectors \citep[e.g.,][]{einstein_telescope,cosmic_explorer} will be able to probe directly past the peak of the merger rate and help investigate the BBH mass distribution across cosmic time.

We also showed that BBH mergers with larger primary masses from GCs contain slightly larger fractions of 2G BHs, assuming BHs are born with zero spins from collapsing stars. About $30\%$ of BBHs with $M_1\gtrsim40\,\msun$ consist of at least one 2G component, while only about $12\%$ with $M_1 \lesssim 20\,\msun$ have 2G components. This is expected since 2G BHs are, in general, more massive than 1G BHs. Despite the increased presence of 2G BHs in the more massive merging BBHs, which also merge at higher redshifts, we observed no redshift-dependent evolution in the effective spins of cluster BBHs. This is due to the overall fraction of BBHs containing 2G BHs remaining relatively constant over a Hubble time.

\begin{acknowledgments}
    We thank Tom Callister, Giacomo Fragione, Marta Reina-Campos, and Aditya Vijaykumar for helpful discussions. We also thank the anonymous referee for helpful comments. C.S.Y. acknowledges support from the Natural Sciences and Engineering Research Council of Canada (NSERC) DIS-2022-568580. M.F. is supported by NSERC RGPIN-2023-05511.
\end{acknowledgments}

\software{\texttt{CMC} \citep{Joshi_2000,Joshi_2001,Fregeau_2003, fregeau2007monte, Chatterjee+2010,Chatterjee+2013b,Umbreit_2012,Morscher+2015,Rodriguez+2016million,CMC1}, \texttt{Fewbody} \citep{fregeau2004stellar}, \texttt{COSMIC} \citep{cosmic}}

\bibliography{bbh_redshift}
\bibliographystyle{aasjournal}

\end{document}